\documentclass[conference,letter]{IEEEtran}
\usepackage{epsfig,graphics,amssymb}
\usepackage{amscd}
\usepackage{amsmath,bbm}
\usepackage{mathtools}
\usepackage{enumerate}
\usepackage{theorem}
\usepackage{cite}
\usepackage[font={small}]{caption}
\usepackage{verbatim}
\usepackage{color}
\theoremstyle{plain} \theorembodyfont{\itshape}
\theoremheaderfont{\scshape}

\theorembodyfont{\itshape} \theoremheaderfont{\scshape}

\theoremstyle{plain} \theorembodyfont{\itshape}
\theoremheaderfont{\scshape}

\newtheorem{proposition}{Proposition}

\usepackage[process=auto]{pstool}
\usepackage{tikz}
\newcommand{\T}{{\rm T}}

\newcommand{\matr}[1]{\mathlette{\boldmath}{#1}}

\newcommand{\tr}{\text{tr}}

\newcommand{\argmin}{$\text{argmin}$}

\def\argmin{\mathop{\rm argmin}}

\def\expect{\mathop{\mbox{$\mathsf{E}$}}}

\newcommand{\e}{{\rm e}}

\newcommand{\Q}{{\rm Q}}

\newcommand{\define}{\stackrel{\triangle}{=}}

\newcommand{\R}{{\rm R}}
\newcommand{\dif}{{\rm d}}
\newcommand{\Dif}{{\rm D}}

\begin{document}

\title{A New Class of Nonlinear Precoders for Hardware Efficient Massive  MIMO Systems}
\author{\IEEEauthorblockN{ Mohammad A.\ Sedaghat, Ali Bereyhi, Ralf~R.~M{\"u}ller }
\IEEEauthorblockA{$^*$ Friedrich-Alexander Universit\"at Erlangen-N\"urnberg, Erlangen, Germany
\IEEEauthorblockA{Emails: \{mohammad.sedaghat, ali.bereyhi, ralf.r.mueller\}@fau.de}
}}

\maketitle

\begin{abstract}
A general class of nonlinear Least Square Error (LSE) precoders in multi-user multiple-input multiple-output systems is analyzed using the replica method from statistical mechanics. A single cell downlink channel with $N$ transmit antennas at the base station and $K$ single-antenna users is considered. The data symbols are assumed to be iid Gaussian and the precoded symbols on each transmit antenna are restricted to be chosen from a predefined set $\mathbbmss{X}$. The set $\mathbbmss{X}$ encloses several well-known constraints in wireless communications including signals with peak power, constant envelope signals and finite constellations such as Phase Shift Keying (PSK). 
 We determine the asymptotic distortion of the LSE precoder under both the Replica Symmetry (RS) and the one step Replica Symmetry Breaking (1-RSB) assumptions. For the case of peak power constraint on each transmit antenna, our analyses under the RS assumption show that the LSE precoder can reduce the peak to average power ratio to 3dB without any significant performance loss. 
For PSK constellations, as $N/K$ grows, the RS assumption fails to predict the performance accurately and therefore, investigations under the 1-RSB assumption are further considered. The results show that the 1-RSB assumption is more accurate.
\end{abstract}

\section{Introduction}

%
Massive Multiple-Input Multiple-Output (MIMO) is among the main technologies for the next generation of wireless networks \cite{andrews2014will,marzetta2010noncooperative}. In massive MIMO systems, multi-antenna base stations utilize precoding techniques to focus the power to desired users. This shifts a large portion of the system's overall processing to base stations which in contrast to power-limited users, do not face a computing power constraint.

So far, several precoding schemes have been proposed including linear and nonlinear schemes. Linear schemes mainly consist of Match Filtering (MF), Zero Forcing (ZF) and Regularized Zero Forcing (RZF), where in practice each of them could be preferred regarding the desired tradeoff between the complexity and performance \cite{rusek2013scaling}. As examples of nonlinear schemes, one names Tomlinson-Harashima \cite{fischer2002space} and vector precoding \cite{hochwald:05}. Regarding data support, precoding schemes can be designed for finite or infinite input alphabets. In finite alphabet cases, the users data symbols are chosen from a finite and countable set, e.g., Phase Shift Keying (PSK) constellation \cite{zeng2012linear}. 

In this paper, we consider a nonlinear Least Square Error (LSE) precoder in which the signal on each transmit antenna is restricted to be chosen from a predefined set. 
The main motivation for investigating the LSE precoder is to model a large variety of signal constraints at massive MIMO base stations which allow us to use more efficient hardware.
As an example, the LSE precoder can be designed to fulfill an instantaneous peak power constraint on each transmit antenna which avoids clipping at power amplifiers. Furthermore, the LSE precoder are able to fix the envelope of transmitted signals to increase the power efficiency of power amplifiers which was initially investigated in \cite{mohammed2013per}. The LSE precoder also enable us to have finite alphabet signals on antennas\footnote{Note that the constraints considered in this paper are on the signals on the antennas and this is different than the case of having constraint on the input data signal of users.}
 which are required in the recently introduced Load Modulated Single-RF (LMSRF) MIMO transmitters \cite{sedaghat2016load}. In LMSRF, the signal on each antenna is taken from a discrete constellation due to using limited number of switches . 

Considering a large number of transmit antennas, we study the LSE precoder in the asymptotic regime. Our performance analyses are based on the replica method introduced in the context of statistical mechanics. The LSE precoder is analyzed in frequency-flat fading channels and it is shown that the same performance is achieved in frequency-selective fading channels utilizing Orthogonal Frequency Division Multiplexing (OFDM). Both the Replica Symmetry (RS) and one-step Replica Symmetry Breaking (1-RSB) assumptions are applied under some known signal constraints such as signals with peak and average power constraints, constant envelope signals and PSK signals. It is shown that in the case of peak power constraint, the RS prediction is consistent with the numerical results, although in some cases it might not be exact. However, the RS assumption does not give an accurate solution for Binary Shift Keying (BPSK) and Quadrature Phase Shift Keying (QPSK) signals on the transmit antennas and 1-RSB improves the prediction in these cases.


\textbf{Notation:} We use bold lowercase letters for vectors and bold uppercase letters for matrices. Conjugate transpose of a matrix is denoted by $\cdot^\dagger$, the transpose itself is shown by $\cdot^{\sf T}$.  Moreover, the complex set is shown by $\mathbbmss{C}$. $F_{\matr{b}}(\cdot)$ denotes the cumulative distribution function (cdf) of $\matr{b}$ and the Kronecker product is shown by $\otimes$. The real and imaginary parts are denoted by $\Re$ and $\Im$, respectively, and $\expect$ represents the mathematical expectation. Furthermore, we define $\Dif z\define \frac1 \pi\e^{-|z|^2}\dif z$ and ${\rm Vec}(\matr{A})$ to be the vector obtained by stacking
the columns of $\matr{A}$.
\section{Problem formulation}
Let's consider the general problem of designing a precoder for a massive MIMO system with $N$ transmit antennas and $K$ single-antenna users. The channel is assumed to be a frequency-flat fading channel. The generalization to the case of frequency-selective fading channels and OFDM signal is presented in Appendix A where we show that the same result holds for the case of frequency-selective channels. The inter-cell interference is neglected and it is assumed that the channel matrix is perfectly known at the base station. 
Let $\matr{u}\in \mathbbmss{C}^K$ and $\matr{H}\in \mathbbmss{C}^{K\times N}$ be the data vector of the users and the channel matrix, respectively, and $\matr{v}\in \mathbbmss{X}^N$ denotes the precoded vector signal where $\mathbbmss{X}$ is a predefined set. The received vector at the user terminals reads
\begin{eqnarray}
\matr{y}=\matr{Hv}+\matr{n},
\end{eqnarray}
where $\matr{y}=[y_1,\cdots,y_K]^\T$ with $y_i$ being the received signal at the $i$th user terminal, and $\matr{n}$ being the zero mean additive white Gaussian noise vector whose elements have the variance of $\sigma_n^2$. We define the non-linear LSE precoder with the following rule
 \begin{eqnarray}\label{precrule}
\matr{v}=\argmin_{\matr{x} \in \mathbbmss{X}^N} \|\matr{Hx}-\sqrt{\gamma}\matr{u}  \|^2+\lambda \|\matr{x}\|^2,
\end{eqnarray}
where $\gamma$ is a positive constant and $\lambda$ is a tuning parameter (Lagrange multiplier) controlling the total transmit power. In case of $\mathbbmss{X}=\mathbbmss{C}$, our nonlinear LSE precoding scheme reduces to the linear scheme
\begin{align}\label{MMSEpr}
\matr{v}=\sqrt{\gamma}\matr{H}^\dagger \left(\matr{H} \matr{H}^\dagger +\lambda \matr{I} \right)^{-1}\matr{u}
\end{align} 
which is known as the regularized zero-forcing precoder \cite{peel2005vector}. The precoding procedure, however, does not take a simple form for a general $\mathbbmss{X}$. As an example, consider the LMSRF MIMO transmitter which chooses a finite constellation with respect to the number of discrete load modulators' states \cite{sedaghat2014novel,sedaghat2016load}. Another example is the case of constant envelope precoding on each antenna \cite{mohammed2013per} where
\begin{eqnarray}
\vert v_i\vert ^2=P \quad \forall i \in\{1,\cdots,N\}.
\end{eqnarray}
Here, the Peak~to~Average Power Ratio~(PAPR)~is~small, around $3$ or $4$ decibels depending the pulse shaping filter. In these cases, the classical tools fail to analyze the optimization problem in \eqref{precrule}. We therefore invoke the replica method developed in statistical mechanics to determine the large system performance of the precoder by calculating the asymptotic distortion defined as
\begin{eqnarray}\label{orgopt}
\mathsf{D} =\lim_{K\uparrow \infty} \frac{1}{K} \expect\|\matr{Hv}-\sqrt{\gamma}\matr{u}  \|^2
\end{eqnarray}
when the inverse load factor defined as $\alpha\define {N}/{K}$, is kept fixed. Our analysis determines the asymptotic distortion defined in \eqref{orgopt} without finding the explicit solution of the optimization problem \eqref{precrule}. This allows us to have an estimate of the best performance in order to have a reference performance measure
for comparing the practical algorithms. 

Throughout the analysis, we set the data symbols of the users to be independent and identically distributed (iid) Gaussian, i.e., $\matr{u}\sim \mathcal{CN}(\matr{0},\sigma_u^2\matr{I})$.

The asymptotic distortion measure $\mathsf{D}$ can be used to derive a lower bound for the ergodic achievable rate of the users in the downlink channel as follows. Let $R_i$ be the ergodic achievable rate of the $i$th user. A lower bound for the ergodic rate $R_i$ is obtained when we impose the worst case scenario for the interference at each user terminal which implies the Gaussian distributed interference at each user terminal. Note that this is true only in the case of Gaussian distributed input signals. Then, we obtain the following bound on the average ergodic rate of the users
\begin{eqnarray}
\frac{1}{K}\sum_{i=1}^K R_{i}\geq \frac{1}{K}\sum_{i=1}^K\expect\limits_{\matr{H}}\log\left(1+\frac{\gamma\sigma_u^2}{\sigma_n^2+I_i(\matr{H})}\right),
\end{eqnarray}
where $I_i(\matr{H})$ is the interference power at the $i$th user terminal. Using Jensen's inequality and the fact that the function $\log\left(1+\frac{\gamma\sigma_u^2}{\sigma_n^2+x}\right)$ is convex, we obtain
\begin{eqnarray}
\frac{1}{K}\sum_{i=1}^K  R_i\geq \log\left(1+\frac{\gamma\sigma_u^2}{\sigma_n^2+\frac{1}{K}\sum\limits_{i=1}^K \expect\limits_{\matr{H}}I_i(\matr{H})}\right).
\end{eqnarray}
It is easy to show that $\frac{1}{K}\sum\limits_{i=1}^K \expect\limits_{\matr{H}}I_i(\matr{H})=\mathsf{D}$ for $K\rightarrow \infty$, and therefore,
\begin{eqnarray}
\frac{1}{K}\sum_{i=1}^K  R_i\geq \log\left(1+\frac{\gamma\sigma_u^2}{\sigma_n^2+\mathsf{D}}\right).
\end{eqnarray}
In the case that users have symmetry, e.g., when the users are uniformly distributed in an area, it is easy to show that the ergodic rate of each user is also larger than $\log\left(1+\frac{\gamma\sigma_u^2}{\sigma_n^2+\mathsf{D}}\right)$.

In this paper we use the replica method to analyze the LSE precoding in a very general case. The replica method also known as the ``replica trick'' is a non-rigorous method developed for asymptotic analysis in statistical mechanics. The method has been rigorously justified in some few cases, e.g., for the system whose matrix has a semicircular eigenvalue distribution. Furthermore, it has been shown that the replica method gives valid predictions for several known problems, and thus it is widely employed for large system analysis of communication systems \cite{guo:04,mueller:08,zaidel2012vector}. 

\section{Large System Analysis}
Define $\matr{R}\define \matr{H}^\dagger \matr{H}$. We start our analysis by determining 
\begin{eqnarray}
&&\breve{\mathsf{D}}\define\lim_{K\uparrow \infty}\frac{1}{K}\expect\min_{\matr{x} \in \mathbbmss{X}^N} \|\matr{Hx}-\sqrt{\gamma}\matr{u}  \|^2 + \lambda \| \matr{x}\|^2 
\end{eqnarray}
which reads 
\begin{align}\label{optinitialrep}
\breve{\mathsf{D}}=\gamma\sigma_u^2+\lim_{K\uparrow \infty}\frac{1}{K}\expect\min\limits_{\matr{x} \in \mathbbmss{X}^N} \mathsf{g}(\matr{x})
\end{align}
with the function $\mathsf{g}(\matr{x})$ being defined as
\begin{eqnarray}
\mathsf{g}(\matr{x})\define\matr{x}^\dagger \matr{R} \ \matr{x}-2\sqrt{\gamma} \Re\left\{\matr{x}^\dagger \matr{H}^\dagger \matr{u}\right\}+ \lambda \matr{x}^\dagger \matr{x}.
\end{eqnarray}
Using Varadhan's theorem, one can write
\begin{align}\label{Farhadkhan}
&\min\limits_{\matr{x}\in \mathbbmss{X}^N}{\mathsf{g}} (\matr{x})
=-\lim\limits_{\beta\uparrow \infty}\frac{1}{\beta}
 \log\sum_{\matr{x} \in \mathbbmss{X}^N}\e^{-\beta{\mathsf{g}} (\matr{x})}.
\end{align}
From \eqref{optinitialrep} and \eqref{Farhadkhan}, the evaluation of $\breve{\mathsf{D}}$ needs a logarithmic expectation to be determined which is not a trivial task. Thus, we use 
\begin{equation}\label{replicacont}
\expect \log(t) =\lim_{n\downarrow 0}\frac{\partial}{\partial n} \log \expect t^n ,
\end{equation}
for some positive random variable $t$. 
Consequently, we have 
\begin{align}
\breve{\mathsf{D}}&= \gamma\sigma_u^2 -\lim_{K,\beta\uparrow \infty}\frac{1}{\beta K}\lim_{n\downarrow 0}\frac{\partial }{\partial n} 
\log \expect \left[\sum_{\matr{x} \in \mathbbmss{X}^N}\e^{-\beta \mathsf{g}(\matr{x})  }\right]^n\nonumber \\ 
&= \gamma\sigma_u^2 -\lim_{\beta\uparrow \infty}\frac{1}{\beta}\lim_{n\downarrow 0}\frac{\partial }{\partial n} \Xi_n,
\end{align}
where $\Xi_n$ denotes the corresponding term in the first line.
Here, the replica method suggests us to consider the replica continuity assumption and do the further analysis as follows: First, determine $\Xi_n$ for an integer $n$, and then, assume that the analytic continuation of $\Xi_n$ onto the real line gives the same result. For details about the validity of this assumption, the reader is referred to \cite{talagrand2000mean}. Thus, we obtain
\begin{align}\label{Xin}
&\Xi_n 
=\lim_{K\uparrow \infty} \frac{1}{K} \log  
 \sum_{\{\matr{x}_a \}} \expect \e^{-\beta  \sum\limits_{a=1}^n \mathsf{g}(\matr{x}_a)    } 
\end{align}
where $\{\matr{x}_a \}$ 
denotes the replicas \mbox{$\{\matr{x}_1,\ldots,\matr{x}_n\} \in \mathbbmss{X}^N\times \ldots \times \mathbbmss{X}^N$}.
Using the independency of $\matr{u}$ and $\matr{H}$, the expectations over $\matr{u}$ and $\matr{H}$ separate, and thus, the summation on the right hand side of \eqref{Xin} reduces to
\begin{align}
 \sum_{\{\matr{x}_a \}} \expect_\matr{H} \e^{-\beta\sum\limits_{a=1}^{n}\left[\matr{x}_a^\dagger \matr{R} \hspace{.5mm} \matr{x}_a+ \lambda \matr{x}_a^\dagger \matr{x}_a\right] +\beta^2\gamma\sigma_u^2 \|\sum\limits_{a=1}^n \matr{H} \matr{x}_a\|^2}
\end{align}
By defining the matrix $\matr{V}$ as 
\begin{eqnarray}\label{Veqn}
\matr{V}\define\frac 1N \left[ \matr{x}_1,\cdots,\matr{x}_n \right] \Gamma \left[ \matr{x}_1,\cdots,\matr{x}_n \right]^\dagger
\end{eqnarray} 
with $\Gamma$ being an $n\times n$ matrix with entries 
\begin{eqnarray}
\zeta_{i,j}\define -\beta\gamma\sigma_u^2 +\delta_{i,j},
\end{eqnarray}
$\Xi_n$ is found as 
\begin{align}\label{beforeHarish}
&\Xi_n =\lim_{K\uparrow \infty}\frac{1}{K} \log  \sum_{\{\matr{x}_a \}}\e^{-\beta \lambda \sum_{a=1}^n\matr{x}_a^\dagger \matr{x}_a}\expect_\matr{H} \e^{-\beta N \tr\left( \matr{R}  \matr V\right)}.
\end{align}

Suppose that the empirical distribution of the eigenvalues of $\matr{R}$ converges to a deterministic distribution, and denote~the corresponding cdf with $F_{\matr{R}}(\lambda)$. The Stieltjes transform of the distribution $F_{\matr{R}}(\lambda)$ is defined as ${\mathrm G}_{\matr{R}}(s)=\expect (\lambda-s)^{-1}$.~The corresponding $\mathrm{R}$-transform is then defined as 
\begin{align}
\R_{\matr{R}}(w)=\mathrm G_{\matr{R}}^{-1}(w)-w^{-1}
\end{align}
where $\mathrm G_{\matr{R}}^{-1}(w)$ denotes the inverse with respect to composition. Noting that the expectation in \eqref{beforeHarish} is a spherical integral, we use the results from \cite{guionnet:05} which state
\begin{eqnarray}
\expect_\matr{H} \e^{-\beta N \tr\left( \matr{R}  \matr V\right)}=\e^{-N \sum\limits_{i=1}^N \int\limits_0^{\beta \tilde{\lambda}_i}\R_{\matr{R}}(-w)\dif w} ,
\end{eqnarray}
as $N \uparrow \infty$ with $\tilde{\lambda}_1,\cdots, \tilde{\lambda}_N$ being the eigenvalues of $\matr{V}$. The matrix $\matr{V}$ has only $n$ nonzero eigenvalues which are equal to the nonzero eigenvalues of 
\begin{eqnarray}
\matr{G}=\frac{1}{N} \ \Gamma  \left[\matr{x}_1,\cdots,\matr{x}_n\right]^\dagger\left[\matr{x}_1,\cdots,\matr{x}_n\right].
\end{eqnarray}
Consequently, denoting the eigenvalues of $\matr{G}$ by $\lambda_1,\cdots, \lambda_n$, 
\begin{eqnarray}
\expect_\matr{H} \e^{-\beta N \tr\left( \matr{R}  \matr V\right)}=\e^{-N\left( \sum\limits_{i=1}^n \int\limits_0^{\beta \lambda_i}\R_{\matr{R}}(-w)\dif w+\epsilon_N \right)},
\end{eqnarray}
where $\epsilon_N$ tends to zero when $N\uparrow \infty$.
In order to find $\Xi_n$, we need to sum over the $Nn$-dimensional space. We determine the summation by taking the same approach as in \cite{mueller:08}. We split the space into the subshells
\begin{eqnarray}
\mathcal{S}(\matr{Q})\define \{\matr{x}_1,\cdots,\matr{x}_n|\matr{x}_a^\dagger \matr{x}_b=NQ_{ab}\}
\end{eqnarray}
where $Q_{ab}$ is the $(a,b)$th entry of the matrix
\begin{eqnarray}
\matr{Q}=\frac{1}{N} \ [\matr{x}_1,\cdots,\matr{x}_n]^\dagger[\matr{x}_1,\cdots,\matr{x}_n].
\end{eqnarray}
Substituting in \eqref{beforeHarish}, $\Xi_n$ reduces to
\begin{align} \label{replIQ}
&\Xi_n 
= \lim_{K\uparrow \infty}\frac{1}{K} \log   \int \e^{N\mathcal{I}(\matr{Q})} \e^{-N\mathcal{G}(\matr{Q})}\mathcal{D}\matr{Q},
\end{align}
where the function $\mathcal{G}(\matr{Q})$ is defined as
\begin{eqnarray}
\mathcal{G}(\matr{Q})\define\beta \lambda \sum_{a=1}^n\frac{\matr{x}_a^\dagger \matr{x}_a}{N}+\sum\limits_{i=1}^n \int\limits_0^{\beta \lambda_i}\R_\matr{R}(-w)\dif w ,
\end{eqnarray}
and $\e^{N\mathcal{I}(\matr{Q})}$ is the Jacobian of the integral; moreover, we define 
\begin{equation}
{\cal D}\matr Q \define \prod\limits_{a=1}^n \prod \limits_{b=a+1}^n {\rm d}{\Re}Q_{ab}{\rm d}{\Im}Q_{ab}.
\end{equation}
In order to calculate the Jacobian term in \eqref{replIQ}, we firstly write
\begin{align}
&{\rm e}^{N\mathcal{I}(\matr Q)} =\int \prod_{a\leq b}
 \delta\left( \Re\left[{\matr{x}}_a^\dagger {\matr{x}}_b-NQ_{ab}\right]  \right) \times
 \nonumber \\ 
 &~~~~~~~~~~~\delta\left( \Im\left[{\matr{x}}_a^\dagger {\matr{x}}_b-N Q_{ab}\right]  \right)\  \prod_{a=1}^{n}{\rm d}F_{\matr x}(\matr{x}_a).
\end{align}
Then, we introduce a new matrix $\tilde{\matr{Q}}$ in the complex frequency domain. Following the lines in \cite[eq.(52)-(58)]{mueller:08} and defining $\mathcal{J}\define(t-j\infty;t+j\infty)$ for some $t\in \mathbbmss{R}$, we obtain
\begin{eqnarray}
\e^{N\mathcal{I}(\matr{Q})}=\int\limits_{\mathcal{J}^{n^2}}\e^{-N\tr[\tilde{\matr{Q}} \matr{Q}]+N\log \mathcal{M } (\tilde{\matr{Q}})} \mathcal{D}\matr{Q}
\end{eqnarray}
where the function $\mathcal{M}(\tilde{\matr{Q}})$ is defined as
\begin{eqnarray}
\mathcal{M}(\tilde{\matr{Q}})\define\sum\limits_{\{x_a\}} \e^{\sum\limits_{a,b}x_a^*x_b\tilde{Q}_{ab}}.
\end{eqnarray}
In the large system limit, the integration in \eqref{replIQ} is dominated by the integrand at the saddle point. In order to calculate the saddle point of the integrand, one needs to impose a structure on the matrices $\matr{Q}$ and $\tilde{\matr{Q}}$. The most primary structure is imposed by the RS assumption. In the RS assumption, it is postulated that the saddle point matrices which dominate the integration in \eqref{replIQ} are invariant to permutation of the replica indices. Therefore, following \cite{mueller:08}, under the RS assumption we set $Q_{a,b}=q$ and $\tilde{Q}_{a,b}=\beta^2 f^2$ for all $a\neq b$ and $Q_{a,a}=q+\chi/\beta$ and $\tilde{Q}_{a,a}=\beta^2f^2-\beta e$ for all $a$
where $\{q,\chi,f,e\}$ are some positive finite constants. Substituting in \eqref{replIQ}, $\Xi_n$ can be analytically calculated, and consequently, $\breve{\mathsf{D}}$ is determined accordingly. It is then straightforward to evaluate the asymptotic distortion $\mathsf{D}$ explicitly. The final expression for the asymptotic distortion under the RS assumption is stated in Proposition~\ref{prop1}.

\begin{proposition}\label{prop1}
\normalfont
Under the RS assumption, the asymptotic distortion converges to
\begin{eqnarray}
\mathsf{D}=\gamma\sigma_u^2+\alpha \hspace{.5mm} \frac{\partial}{\partial \chi}\left[ (q-\chi\gamma\sigma^2_u)\chi \R_{\matr{R}}(-\chi)\right],
\end{eqnarray}
as $K,N\uparrow \infty$ and the inverse load factor $\alpha$ is kept fixed. $q$ and $\chi$ are solutions to the following two coupled equations
{\small \begin{align}
&\chi=\frac 1f \Re \int_{\mathbbmss{C}} \argmin\limits_{x\in \mathbbmss{X}} \left|z-\frac{\R_{\matr{R}}(-\chi)+\lambda }{f}x  \right|z^* \Dif z   \\
&q=\int_{\mathbbmss{C}} \left|\argmin\limits_{x\in \mathbbmss{X}} \left|z-\frac{\R_{\matr{R}}(-\chi)+\lambda}{f} x \right| \right|^2 \Dif z
\end{align}}
where 
\begin{eqnarray}
f\define \sqrt{(q-\chi \gamma\sigma_u^2)\R^\prime_\matr{R}(-\chi)+\gamma\sigma^2_u \R_{\matr{R}}(-\chi)}.
\end{eqnarray}
\hfill $\blacksquare$
 \end{proposition}


 Although the replica symmetry assumption has given the exact solution for some problems \cite{tanaka:02,montanari2006analysis}, there are several examples in which this assumption fails to give valid prediction of the solution \cite{zaidel2012vector}. For these cases in order to have a more precise prediction, one needs to employ the $r$-step RSB assumption which imposes a more generalized structure on the matrices $\matr{Q}$ and $\tilde{\matr{Q}} $. Here, we consider the 1-RSB assumption which postulates \cite{zaidel2012vector}
\begin{align} \label{RSBstr}
\matr{Q}&=q_1 \matr{1}_{n}+p_1\matr{I}_{\frac{n\beta}{\mu_1}} \otimes \matr{1}_{\frac{\mu_1}{\beta}}+\frac{\chi_1}{\beta} \matr{I}_{n}, \\
\tilde{\matr{Q}}&=\beta^2 f_1^2 \matr{1}_{n} + \beta^2 g_1^2\matr{I}_{\frac{n\beta}{\mu_1}} \otimes \matr{1}_{\frac{\mu_1}{\beta}} -\beta e_1 \matr{I}_n,
\end{align}
where $q_1,p_1,\chi_1,\mu_1,f_1,g_1,e_1$ are non-negative real numbers, $\matr{1}_{n}$ is an $n\times n$ matrix with all elements equal to $1$ and $\matr{I}_n$ is the $n\times n$ identity matrix. 
With same steps as for Proposition~\ref{prop1}, the asymptotic distortion is determined under the 1-RSB assumption as in the following proposition.

 \begin{proposition}\label{prop2}
 \normalfont
Under the 1-RSB assumption, the asymptotic distortion converges to
 {\small\begin{align}
\mathsf{D}=&\gamma\sigma_u^2-\frac{\alpha\chi_1}{\mu_1}
\R_{\matr{R}}(-\chi_1)
+ \alpha\left[ q_1+\frac{\eta_1}{\mu_1}-2\gamma\sigma_u^2\eta_1\right]\R_{\matr{R}}(-\eta_1) \nonumber\\
&-\alpha\eta_1\left[q_1-\gamma\sigma_u^2\eta_1\right]\R_{\matr{R}}^{\prime}(-\eta_1),
\end{align}}
 as $K,N\uparrow \infty$ and the inverse load factor $\alpha$ is kept fixed. The set of scalars $\{q_1,p_1,\chi_1,\mu_1\}$ is calculated through the coupled equations
{\small\begin{align}
&\eta_1=
\frac{1}{f_1}\int \int\Re\left\{z^*\argmin|f_1z+g_1y-e_1x |   \right\}\tilde{\mathcal{Y}}(y,z)\Dif z \Dif y,\nonumber \\
&q_1+p_1=\int \int\left|\argmin|f_1z+g_1y-e_1x |   \right|^2\tilde{\mathcal{Y}}(y,z)\Dif z \Dif y, \nonumber \\
&\eta_1+\mu_1q_1= \nonumber \\
&\hspace{3mm}\frac{1}{g_1}\int \int\Re\left\{y^*\argmin|f_1z+g_1y-e_1x |   \right\}\tilde{\mathcal{Y}}(y,z)\Dif z \Dif y,
\end{align}}
and
{\small\begin{align}
\int_{\chi_1}^{\eta_1} &\R_{\matr{R}}(-w)\dif w=\int \log \int \mathcal{Y}(y,z)\Dif y \Dif z-2\chi_1\R_{\matr{R}}(-\chi_1)\nonumber \\
&+(\mu_1q_1+2\eta_1-2\mu_1 \eta_1 \gamma\sigma_u^2-2\chi_1\mu_1\gamma\sigma_u^2)\R_{\matr{R}}(-\eta_1)\nonumber \\
&-2\mu_1\eta_1(q_1-\gamma\sigma_u^2\eta_1)	\R_{\matr{R}}^{\prime}(-\eta_1) 
+ \lambda\mu_1(p_1+q_1),
\end{align}}
where $\eta_1=\chi_1+\mu_1p_1 $,
\begin{eqnarray}
\mathcal{Y}(y,z)=\e^{-\mu_1 \min\limits_{x\in \mathbbmss{X}}e_1 |x|^2-2\Re\{x(f_1z^*+g_1y^*)\}},
\end{eqnarray}
and the function $\tilde{\mathcal{Y}}(y,z)$ being defined as
\begin{eqnarray}
\tilde{\mathcal{Y}}(y,z)=\frac{\mathcal{Y}(y,z)}{\int_{\mathbbmss{C}}  \mathcal{Y}(\tilde{y},z)\Dif \tilde{y}}.
\end{eqnarray}
Moreover, the parameters $\{e_1,f_1,g_1\}$ are determined as
{\begin{align}
&e_1=\R_{\matr{R}}(-\chi_1)+\lambda\nonumber \\
&f_1=\sqrt{\gamma\sigma_u^2\R_{\matr{R}}(-\eta_1) + (q_1-\gamma\sigma_u^2\eta_1)\R_{\matr{R}}^{\prime}(-\eta_1)} \nonumber \\
&g_1=\sqrt{\frac{\R_{\matr{R}}(-\chi_1)-\R_{\matr{R}}(-\eta_1)}{\mu_1}}.
\end{align}}
\hfill $\blacksquare$

\end{proposition}

Note that letting $p_1=0$ reduces the 1-RSB solution to RS. This means that one of the 1-RSB solutions is always the RS solution. 
In fact, the coupled equations in both the RS and 1-RSB cases may have multiple solutions in which one of them is the valid saddle point of \eqref{replIQ}. In this case, the saddle point is the solution which minimizes a function corresponding to the system, known as the free energy. It can be shown that the free energy of the nonlinear LSE precoder is $-\breve{\mathsf{D}}$. This means that the saddle point is the solution which maximizes $\breve{\mathsf{D}}$.
The explicit expression for $\breve{\mathsf{D}}$ can be found in terms of the replica parameters and is skipped here due to lack of space.

\section{Numerical results}
In this section, we numerically investigate several examples of the nonlinear LSE precoder.
Throughout our investigations, we consider $\sigma_u^2=1$ and a channel matrix whose entries are iid with variance $1/N$. Note that Propositions~\ref{prop1} and \ref{prop2} are given for a general channel matrix and are not restricted to the iid case. In Appendix A, we show that the results can be generalized to the case of frequency-selective channels and OFDM signal. For the sake of simplicity, we assume that all the users have the same path loss and consider the extension of the analysis for more general setups in the extended version of the paper. For an iid matrix, we have \cite{mueller:08}
\begin{eqnarray}
\R_{\matr{R}}(w)=\frac{\alpha^{-1}}{1- w}.
\end{eqnarray}
In the following, we consider two examples of massive MIMO systems with input constraint; namely transmitters with peak power constraint on each antenna and systems with PSK constellation on the transmit antennas. We determine the performance of the nonlinear LSE precoder using our replica results.
\subsection{Per-antenna peak and total average power constraints}

\hspace{-.5mm}Consider a setup with constraints on the instantaneous power on each transmit antenna and total average power.  The average power is set by choosing $\lambda$ properly. In this case, $\mathbbmss{X}$ reads
\begin{eqnarray}\label{peakpowerset}
\mathbbmss{X}=\left\{x=r\e^{j\theta}\big|\theta\in[0,2\pi], 0\leq r\leq \sqrt{P}\right\}
\end{eqnarray}
where $P$ is the instantaneous peak power.
We invoke the RS solution for investigation. Using Proposition~1, it can be shown that the parameter $q$ in the RS solution represents the average power per antenna. Moreover, the RS coupled equations become
\begin{align}\label{peakceq1}
\chi&=\sqrt{\frac{\alpha}{q+\gamma\sigma_u^2}} \left( 1+\chi \right) h, \nonumber \\
q&=c^2\left[ 1-\e^{-\frac{P}{c^2}} \right] 
\end{align}
for $h$ being defined as
\begin{eqnarray}\label{peakceq2}
h&=& c-c\e^{-\frac{P}{c^2}}+\sqrt{P \pi} \ {\Q}\left(\frac{\sqrt{2P}}{c} \right),
\end{eqnarray}
and $c$ denoted by
\begin{eqnarray}
c=\frac{\sqrt{\alpha(q+\gamma\sigma_u^2)}}{\alpha \lambda (1+\chi)+1}.
\end{eqnarray}
Furthermore, the asymptotic distortion is determined as 
\begin{eqnarray} \label{avedistpeak}
\mathsf{D}= \frac{q+\gamma\sigma^2_u }{(1+\chi)^2}  .
\end{eqnarray}

\begin{figure}[t!]
\centering
\resizebox{1\linewidth}{!}{
\pstool[width=0.8\linewidth]{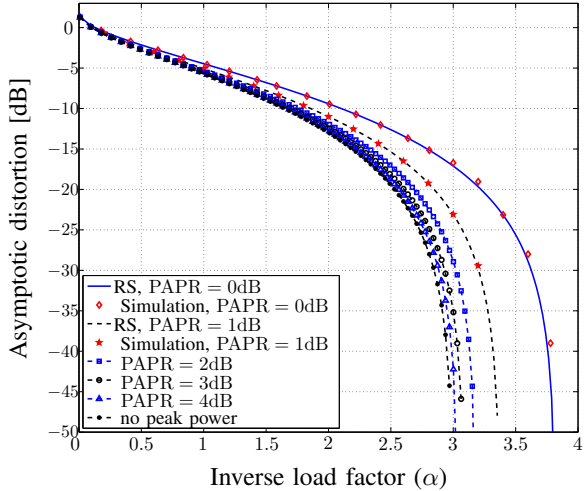}{
\psfrag{average0000000000000000000}[l][l][.6]{$\text{RS, }{\rm PAPR}=0{\rm dB}$}
\psfrag{average1111111100000000000}[l][l][.6]{$~\text{Simulation, }{\rm PAPR}=0{\rm dB}$}
\psfrag{average2222222200000000000}[l][l][.6]{$\text{RS, }{\rm PAPR}=1{\rm dB}$}
\psfrag{average3333333300000000000}[l][l][.6]{$~\text{Simulation, }{\rm PAPR}=1{\rm dB}$}
\psfrag{average4444444400000000000}[l][l][.6]{$~{\rm PAPR}=2{\rm dB}$}
\psfrag{average5555555500000000000}[l][l][.6]{$~{\rm PAPR}=3{\rm dB}$}
\psfrag{average6666666600000000000}[l][l][.6]{$~{\rm PAPR}=4{\rm dB}$}
\psfrag{average7777777700000000000}[l][l][.6]{ no peak power}
\psfrag{Asymptoticdistortion}[c][c][.8]{Asymptotic distortion [dB]}
\psfrag{alpha}[c][c][.8]{Inverse load factor ($\alpha$)}

\psfrag{0}[c][l][.6]{$0$}
\psfrag{0.5}[c][c][.6]{$0.5$}
\psfrag{1}[c][c][.6]{$1$}
\psfrag{1.5}[c][c][.6]{$1.5$}
\psfrag{2}[c][c][.6]{$2$}
\psfrag{2.5}[c][c][.6]{$2.5$}
\psfrag{3}[c][c][.6]{$3$}
\psfrag{3.5}[c][c][.6]{$3.5$}
\psfrag{4}[c][c][.6]{$4$}

\psfrag{-5}[c][c][.55]{$-5$}
\psfrag{-10}[c][c][.55]{$-10$}
\psfrag{-15}[c][c][.55]{$-15$}
\psfrag{-20}[c][c][.55]{$-20$}
\psfrag{-25}[c][c][.55]{$-25$}
\psfrag{-30}[c][c][.55]{$-30$}
\psfrag{-35}[c][c][.55]{$-35$}
\psfrag{-40}[c][c][.55]{$-40$}
\psfrag{-45}[c][c][.55]{$-45$}
\psfrag{-50}[c][c][.55]{$-50$}

}}
\caption{Asymptotic distortion versus the inverse load factor, i.e., $\alpha=N/K$, for different peak power values on each transmit antenna when the per-antenna average power is set to $0.5$.}

\label{peak_average_fix_q_difflam}
\end{figure}

Fig.~\ref{peak_average_fix_q_difflam} represents the asymptotic distortion versus the inverse load factor for a fixed total average power $q=0.5$, different PAPR defined as $P/q$ and $\gamma=1$. To validate the results by the replica method, we have also plotted some simulation results obtained by CVX, a package for specifying and solving convex programs \cite{cvx}, for $K=200$. It is observed that the analytical results are consistent with the simulation results although RS may not be exact in some cases. Furthermore, for PAPRs equal to or more than $3\ {\rm dB}$, the asymptotic distortion is sufficiently close to the case without peak power constraint. 

\begin{figure}[t]
\centering
\resizebox{1\linewidth}{!}{
\pstool[width=0.8\linewidth]{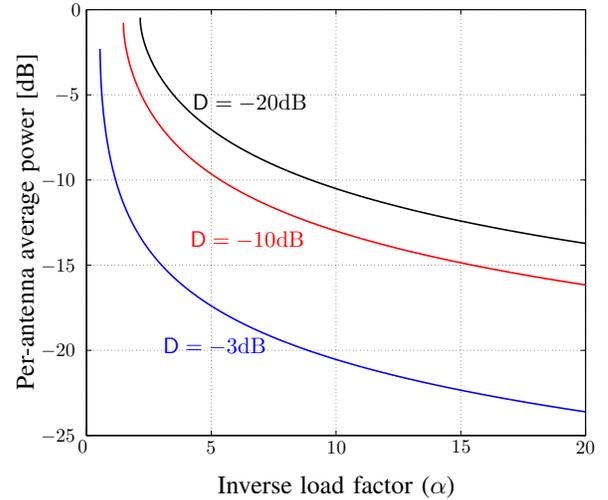}{
\psfrag{Per-antenna average power (dB)}[c][c][.8]{Per-antenna average power [dB]}
\psfrag{D=-20dB}[l][l][.7]{$\mathsf{D}=-20{\rm dB}$}
\psfrag{D=-10dB}[r][r][.7]{{\color{red}$\mathsf{D}=-10{\rm dB}$}}
\psfrag{D=-3dB}[r][r][.7]{{\color{blue}$\mathsf{D}=-3{\rm dB}$}}
\psfrag{alpha}[c][c][.8]{Inverse load factor ($\alpha$)}

\psfrag{0}[c][l][.6]{$0$}
\psfrag{5}[c][c][.6]{$5$}
\psfrag{10}[c][c][.6]{$10$}
\psfrag{15}[c][c][.6]{$15$}
\psfrag{20}[c][c][.6]{$20$}
\psfrag{-25}[c][l][.6]{$-25$}
\psfrag{-5}[c][l][.6]{$-5$}
\psfrag{-10}[c][l][.6]{$-10$}
\psfrag{-15}[c][l][.6]{$-15$}
\psfrag{-20}[c][l][.6]{$-20$}

}}
\caption{Required per-antenna average power versus the inverse load factor, i.e., $\alpha=N/K$, for different asymptotic distortions and $P=1$.}
\label{average_vs_alpha_fixedD}
\end{figure}

 In order to describe the variation of the required average power for a fixed asymptotic distortion with respect to the number of transmit antennas, we consider the case with unit per-antenna peak power constraint and plot the average power per-antenna for given asymptotic distortions. The parameter $\gamma$ is set to $1$. The results are shown in Fig.~\ref{average_vs_alpha_fixedD}. It is observed that the per-antenna average power decays by increasing $\alpha$. By numerical curve fitting, it can be observed that the per-antenna average power converges to $c\alpha ^{\kappa}$ for some constants $c$ and $\kappa=-1$ as $\alpha$ grows unbounded. For bounded $\alpha$, however, $\kappa<-1$. For massive MIMO systems, i.e., $\alpha\gg 1$, with average power constraint, it has been shown that when the base station has perfect channel state information, signal to interference plus noise ratio can be improved by a factor of $\alpha$, asymptotically \cite{rusek2013scaling} which agrees the result given here for the peak power constraint.

%

\begin{figure}[t]
\centering
\resizebox{1\linewidth}{!}{
\pstool[width=0.8\linewidth]{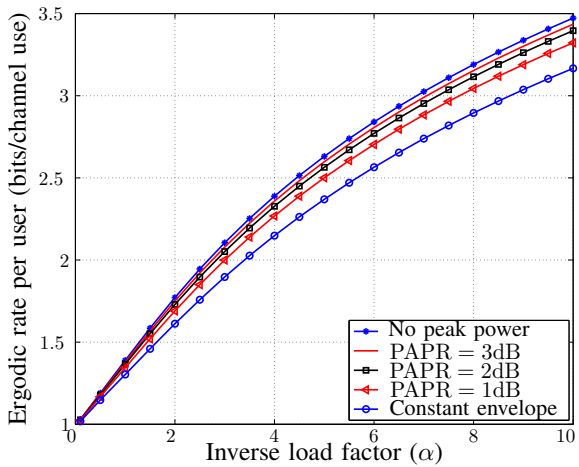}{
\psfrag{Achievable rate (bits/channel use)}[c][c][.8]{Ergodic rate per user (bits/channel use)}
\psfrag{AAAAAAAAAAAAA}[l][l][.7]{~No peak power }
\psfrag{BBBBBBBBBBBBB}[l][l][.7]{ ${\rm PAPR}=3{\rm dB}$}
\psfrag{CCCCCCCCCCCC}[l][l][.7]{ ${\rm PAPR}=2{\rm dB}$}
\psfrag{alpha}[c][c][.8]{Inverse load factor ($\alpha$)}
\psfrag{DDDDDDDDDDDDDD}[l][l][.7]{ ${\rm PAPR}=1{\rm dB}$}
\psfrag{EEEEEEEEEEEEEEEEE}[l][l][.7]{~Constant envelope}

\psfrag{0}[c][l][.6]{$0$}
\psfrag{2}[c][c][.6]{$2$}
\psfrag{4}[c][c][.6]{$4$}
\psfrag{6}[c][c][.6]{$6$}
\psfrag{8}[c][c][.6]{$8$}
\psfrag{10}[c][l][.6]{$10$}
\psfrag{1}[c][l][.6]{$1$}
\psfrag{1.5}[c][l][.6]{$1.5$}
\psfrag{2}[c][l][.6]{$2$}
\psfrag{2.5}[c][l][.6]{$2.5$}
\psfrag{3}[c][l][.6]{$3$}
\psfrag{3.5}[c][l][.6]{$3.5$}

}}
\caption{The lower bound for the ergodic rate per user versus the inverse load factor for different PAPR when $q=1$, $\sigma_n^2=1$ and $\gamma$ is optimized.}
\label{rate}
\end{figure}

Next, the lower bound for the ergodic rate per user is investigated. The noise variance $\sigma_n^2$ and the average transmitted power $q$ are set to 1. An important parameter in this case is $\gamma$. Increasing $\gamma$ outperforms the received SNR but at the same time it also affects the power of interference. The numerical results show that there is an optimum value for $\gamma$ for every $\alpha$ and the average transmitted power $q$. In Fig.~\ref{rate}, the lower bound for the ergodic rate per user is plotted versus the inverse load factor for different peak to average power constraints when the rate is optimized over $\gamma$. Note that although the replica method also predicts the results for $\alpha<1$, but the valid system assumption here is $\alpha\geq1$ since the number of base station antennas should be larger than the number of users. The rate for different PAPRs are quit close. At around $\alpha=5$, for the case of constant envelope signal we need about $20\%$ more antennas to obtain the same performance as in the case of no peak power constraint. Further simulations for the case of constant envelope signal, which are not presented here due to space limitation, show that for $\alpha=5$, about $1.3{\rm dB}$ more transmit power is required to get the same performance compared to the case of no peak power constraint.

\subsection{M-PSK signals on antennas}
\begin{figure}[t!]
\centering
\resizebox{1\linewidth}{!}{
\pstool[width=.88\linewidth]{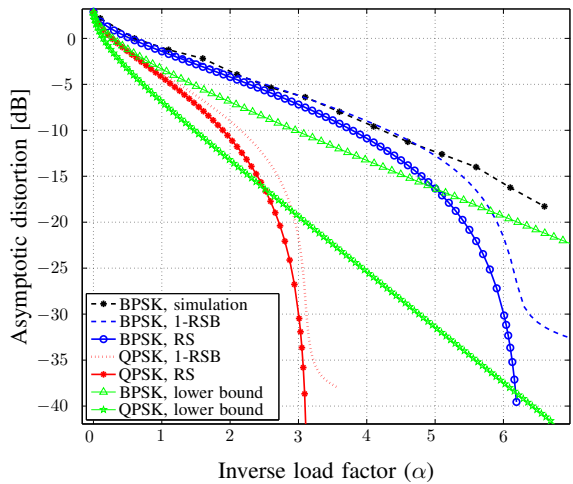}{
\psfrag{Average distortion (dB)}[c][c][.8]{Asymptotic distortion [dB]}
\psfrag{alpha}[c][c][.8]{Inverse load factor ($\alpha$)}
\psfrag{BPSK, simulation}[l][l][.6]{BPSK, simulation}
\psfrag{BPSK, 1-RSB}[l][l][.6]{BPSK, 1-RSB}
\psfrag{BPSK, RS}[l][l][.6]{BPSK, RS}
\psfrag{QPSK, 1-RSB}[l][l][.6]{QPSK, 1-RSB}
\psfrag{QPSK, RS}[l][l][.6]{QPSK, RS}
\psfrag{BPSK, lower boundAAA}[l][l][.6]{BPSK, lower bound}
\psfrag{QPSK, lower bound}[l][l][.6]{QPSK, lower bound}

\psfrag{0}[c][l][.6]{$0$}
\psfrag{5}[c][l][.6]{$5$}
\psfrag{10}[c][l][.6]{$10$}
\psfrag{15}[c][l][.6]{$15$}
\psfrag{20}[c][l][.6]{$20$}
\psfrag{-25}[c][l][.6]{$-25$}
\psfrag{-5}[c][l][.6]{$-5$}
\psfrag{-10}[c][l][.6]{$-10$}
\psfrag{-15}[c][l][.6]{$-15$}
\psfrag{-20}[c][l][.6]{$-20$}
\psfrag{-25}[c][l][.6]{$-25$}
\psfrag{-30}[c][l][.6]{$-30$}
\psfrag{-35}[c][l][.6]{$-35$}
\psfrag{-40}[c][l][.6]{$-40$}
\psfrag{-45}[c][l][.6]{$-45$}
\psfrag{-50}[c][l][.6]{$-50$}

\psfrag{1}[c][c][.6]{$1$}
\psfrag{2}[c][c][.6]{$2$}
\psfrag{3}[c][c][.6]{$3$}
\psfrag{4}[c][c][.6]{$4$}
\psfrag{9}[c][c][.6]{$5$}
\psfrag{6}[c][c][.6]{$6$}
\psfrag{7}[c][c][.6]{$7$}
\psfrag{8}[c][c][.6]{$8$}

}}
\caption{Asymptotic distortion versus the inverse load factor, i.e., $\alpha=N/K$, for BPSK and QPSK constellations.}
\label{psksigma1}
\end{figure}

Let's consider the case of which the signal on each transmit antenna is selected from $M$-PSK constellation. In this case, we have
\begin{eqnarray}
\mathbbmss{X}=\left\{\e^{\hspace{.3mm} jk\frac{2\pi}{M}}\big| k=1,\ldots, M\right\}.
\end{eqnarray}
The constant envelope constraint is obtained easily by letting $M\uparrow \infty$. 
Under the RS assumption, the unit per-antenna average power results in $q=1$ and the parameter $\chi$ reads
\begin{eqnarray}
\chi^{-1}&=&\frac{2}{M \sin(\pi/M)}\sqrt{\pi \frac{1+\gamma\sigma^2_u}{ \alpha}}-1.
\end{eqnarray}
Then, the RS prediction for the asymptotic distortion is 
\begin{eqnarray}\label{PSKdistor}
\mathsf{D}= \frac{1+\gamma\sigma^2_u }{(1+\chi)^2}  .
\end{eqnarray}
%
 Fig.~\ref{psksigma1} shows the asymptotic distortion for BPSK and QPSK constellations. For the sake of comparison, a lower bound for the asymptotic distortion is also plotted\footnote{Due to lack of space, derivation of the lower bound is omitted here. }. For a BPSK constellation, the simulation results using an integer programming algorithm is also plotted considering $N=100$.

For the BPSK case, it is observed that the RS prediction starts to deviate from the simulation results as $\alpha$ increases. The RS prediction even violates the lower bound for $\alpha\geq 5$. This observation clarifies the failure of the RS assumption in this case. To have a better approximation of the exact solution, we have also plotted the 1-RSB prediction in Fig.~\ref{psksigma1}.
The 1-RSB prediction meets RS for small $\alpha$ and deviates as $\alpha$ grows. However, the 1-RSB prediction also fails to approximate the simulation results and violates the lower bound for large $\alpha$. This observation brings this conjecture into mind that the precise approximation of the asymptotic distortion is given by the infinite number of replica breaking steps. Similar results are observed for the QPSK constellation.

\begin{figure}[t]
\centering
\resizebox{1\linewidth}{!}{
\pstool[width=0.8\linewidth]{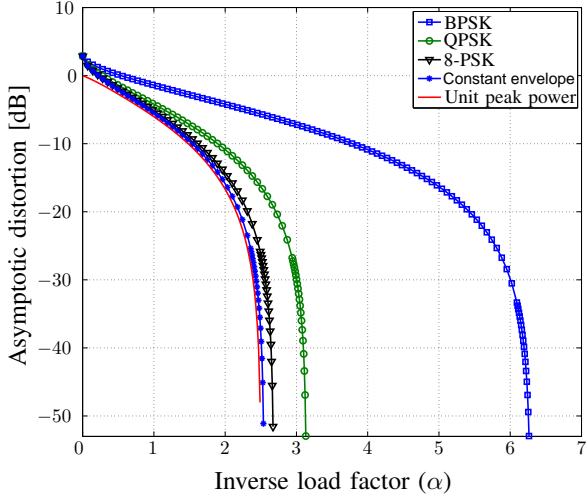}{
\psfrag{BPSK}[l][l][.6]{BPSK}
\psfrag{QPSK}[l][l][.6]{QPSK}
\psfrag{8-PSK}[l][l][.6]{8-PSK}
\psfrag{Constant envelopeAAA}[l][l][.6]{Constant envelope}
\psfrag{Peak power (P=1)}[l][l][.6]{Unit peak power}
\psfrag{Average Distortion (dB)}[c][c][.8]{Asymptotic distortion [dB]}

\psfrag{alpha}[c][c][.8]{Inverse load factor ($\alpha$)}

\psfrag{0}[c][l][.6]{$0$}
\psfrag{10}[c][l][.6]{$10$}
\psfrag{-10}[c][l][.6]{$-10$}
\psfrag{-20}[c][l][.6]{$-20$}
\psfrag{-30}[c][l][.6]{$-30$}
\psfrag{-40}[c][l][.6]{$-40$}
\psfrag{-50}[c][l][.6]{$-50$}

\psfrag{1}[c][c][.6]{$1$}
\psfrag{2}[c][c][.6]{$2$}
\psfrag{3}[c][c][.6]{$3$}
\psfrag{4}[c][c][.6]{$4$}
\psfrag{5}[c][c][.6]{$5$}
\psfrag{6}[c][c][.6]{$6$}
\psfrag{7}[c][c][.6]{$7$}
\psfrag{8}[c][c][.6]{$8$}

}}
\caption{Asymptotic distortion versus the inverse load factor, i.e., $\alpha=N/K$, for BPSK, QPSK and 8-PSK, constant envelope and peak power constraint cases.}
\label{PSK_opt}
\end{figure}

Fig.~\ref{PSK_opt} illustrates the RS prediction for PSK constellations. 
The case of constant envelope signal is also plotted by letting $M\uparrow \infty$. For sake of comparison, we have shown the result for the unit peak power constraint as well. For $M\geq 8$, the results for $M$-PSK and the constant envelope constellations are sufficiently close. Furthermore, the constant envelope constellation and the unit peak power constraint give almost the same asymptotic distortion under the RS assumption.

\section{Conclusions}
The asymptotic performance of the nonlinear LSE precoder was analyzed using the replica method. Under the RS assumption, the asymptotic distortion of the precoder takes a simple form. Based on the investigations, the RS assumption seems to give a valid approximation of the exact solution in the case of a peak power constraint on each antenna and constant envelope signals. For the case with peak power constraint, the numerical results show that the transmit signals with PAPR of about $3{\rm dB}$ perform sufficiently close to the case without peak power constraint. This plays a very important role in practice where low PAPR signals at the transmitter enable us to employ highly efficient nonlinear power amplifiers. 

The RS prediction for an $M$-PSK constellation, however, violates the theoretically rigorous bounds as the inverse load factor $\alpha$, defined as the number of transmit antennas to the number of users, increases. This implies that further exploration based on RSB assumptions is necessary for accurately approximating the performance. We considered the 1-RSB assumption and observed that the solution predicts the simulation results for a larger interval of $\alpha$. 
The results show that for the first time even 1-RSB can be unreliable in wireless communications and more RSB steps need to be considered.



\begin{appendices}
\section{Genealization to frequency-selective fading channels}
Let $L$ be the number of subcarriers and also assume that the channel is frequency-flat at each frequency sub-band. Furthermore, let $\matr{H}_j$ be the channel matrix at $j$th sub-band. The data input vector at the $j$th subcarrier is $\matr{u}_j$. The LSE precoder in this case determines $L$ vectors $\matr{v}_1,\cdots,\matr{v}_L$ to be given to the Inverse Fast Fourier Transform (IFFT) blocks as inputs. Let $\matr{W}$ be the IFFT matrix and $\matr{v}_{\rm t}\define{\rm Vec}\left([\matr{v}_1,\ldots,\matr{v}_L]^\T\right)$. 
The LSE precoder rule is 
\begin{eqnarray}\label{precfft}
\matr{v}_{\rm t}=\argmin\limits_{\matr{W}_{\rm t}\matr{x}_{\rm t}\in \mathbbmss{X}^{NL}} \|\matr{H}_{\rm t}\matr{x}_{\rm t}-\matr{u}_{\rm t}  \|^2,
\end{eqnarray}
where $\matr{H}_{\rm t}$ is a $KL\times NL$ matrix whose $k$th part of its $(i-1)L+k$ columns is the $i$th column of the $\matr{H}_k$ and the remained entries are zero,
$\matr{u}_{\rm t}\define[\matr{u}_1^\T,\ldots,\matr{u}_L^\T]^\T$ and $\matr{W}_{\rm t}$ is an $LN\times LN$ block-diagonal matrix whose $L\times L$ diagonal blocks are equal to $\matr{W}$.

\begin{figure}[t]
\centering
\resizebox{1\linewidth}{!}{
\pstool[width=0.8\linewidth]{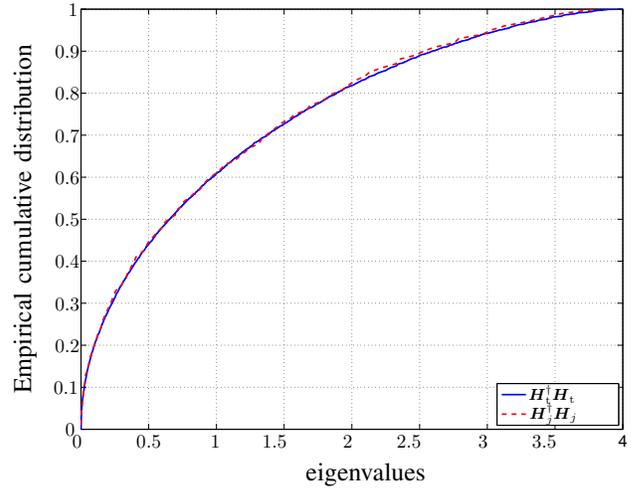}{
\psfrag{Epricial cumulative distribution}[c][c][.8]{Empirical cumulative distribution}
\psfrag{eigenvalues}[l][l][.8]{eigenvalues}
\psfrag{AAAAAAAA}[l][l][.5]{$\matr{H}_{\rm t}^{\dagger}\matr{H}_{\rm t}$}
\psfrag{BBBBBBBB}[l][l][.5]{$\matr{H}_{ j}^{\dagger}\matr{H}_{ j}$}

\psfrag{0}[c][l][.6]{$0$}
\psfrag{0.5}[c][l][.6]{$0.5$}
\psfrag{1}[c][l][.6]{$1$}
\psfrag{1.5}[c][l][.6]{$1.5$}
\psfrag{2}[c][l][.6]{$2$}
\psfrag{2.5}[c][l][.6]{$2.5$}
\psfrag{3}[c][l][.6]{$3$}
\psfrag{3.5}[c][l][.6]{$3.5$}
\psfrag{0.1}[c][c][.6]{$0.1$}
\psfrag{0.2}[c][c][.6]{$0.2$}
\psfrag{0.3}[c][c][.6]{$0.3$}
\psfrag{0.4}[c][c][.6]{$0.4$}
\psfrag{0.5}[c][c][.6]{$0.5$}
\psfrag{0.6}[c][c][.6]{$0.6$}
\psfrag{0.7}[c][c][.6]{$0.7$}
\psfrag{0.8}[c][c][.6]{$0.8$}
\psfrag{0.9}[c][c][.6]{$0.9$}
\psfrag{1}[c][c][.6]{$1$}
}}
\caption{Empirical cumulative distribution of two matrices $\matr{H}_{\rm t}^{\dagger}\matr{H}_{\rm t}$ and $\matr{H}_{ j}^{\dagger}\matr{H}_{ j}$ for $L=32$ and $N=K=100$.}
\label{eigen}
\end{figure}

One can reformulate \eqref{precfft} as 
\begin{eqnarray}
\matr{v}_{\rm t}=\argmin\limits_{\matr{z}_{\rm t}\in \mathbbmss{X}^{NL}} \|\matr{H}_{\rm t}\matr{W}_{\rm t}^\dagger \matr{z}_{\rm t}-\matr{u}_{\rm t}  \|^2,
\end{eqnarray}
using the fact that $\matr{W}_{\rm t}\matr{W}_{\rm t}^\dagger=\matr{I}$.
One can consider an equivalent frequency-flat fading channel with the channel matrix equal to $\matr{H}_{\rm t}\matr{W}_{\rm t}^\dagger$. For the case that $\matr{H}_i$s are iid Gaussian matrices, Fig.~\ref{eigen} compares the empirical cumulative distribution of the eigenvalues of $\matr{R}_{\rm t}=\matr{H}_{\rm t}^\dagger \matr{H}_{\rm t}$ and $\matr{R}_{j}=\matr{H}_{j}^\dagger \matr{H}_{j}$ numerically for $L=32$ and $K=N=100$. It is observed that the both cases have the same distribution. Since the result derived by the replica method depends only on the eigenvalue distribution of $\matr{R}_{\rm t}$, this proves that the LSE precoder in this case has the same performance as in the case of frequency-flat fading channel.

\end{appendices}

\bibliographystyle{IEEEtran} 
\bibliography{lit}

\end{document}